\documentclass[12pt]{revtex4}
\def\comment#1{}

\usepackage{graphicx}
\begin{document}
\title{Neutral nuclear core vs super charged one}
\author{M. Rotondo, R. Ruffini and S.-S. Xue}
\affiliation{ICRA and Physics Department,
University of Rome ``La Sapienz'',
P.le A. Moro 5,
00185 Rome, Italy\\
ICRANeT Piazzale della Repubblica, 10 -65122, Pescara, Italy}

\begin{abstract}
Based on the Thomas-Fermi approach, we describe and distinguish the electron distributions around extended 
nuclear cores: (i) in the case 
that cores are neutral for electrons bound by protons inside cores and proton and electron numbers are the same; 
(ii) in the case that super charged cores are bare, electrons (positrons) produced by 
vacuum polarization are bound by (fly into) cores (infinity).     

\end{abstract}
\maketitle

\vskip0.2cm
\noindent{\it Equilibrium of electron distribution in neutral cores.}\hskip0.3cm     
In Refs.~\cite{ruffinistella80,ruffinistella81,rrx2006}, the Thomas-Fermi approach was used to study the 
electrostatic equilibrium of electron distributions 
$n_e(r)$ around extended nuclear cores, where total proton and electron numbers are the same $N_p=N_e$.
Proton's density $n_p(r)$ is constant inside core $r \le R_c$ and vanishes outside the core $r > R_c$,
\begin{eqnarray}
n_p(r) = n_p\theta(R_c-r),  
\label{pn}
\end{eqnarray}
where $R_c$ is the core radius and $n_p$ proton density. Degenerate electron density,
\begin{eqnarray}
n_e (r)= \frac {1}{3\pi^2\hbar^3}(P_e^F)^3,  
\label{en}
\end{eqnarray}     
where electron Fermi momentum $P_e^F$, Fermi-energy ${\mathcal E}_e(P_e^F)$ and Coulomb potential energy $V_{\rm coul}(r)$
are related by,
\begin{eqnarray}
{\mathcal E}_e(P_e^F) =  [(P_e^Fc)^2+m_e^2c^4]^{1/2}-m_ec^2 - V_{\rm coul}(r).  
\label{eeq0}
\end{eqnarray}
The electrostatic equilibrium of electron distributions is determined by
\begin{eqnarray}
{\mathcal E}_e(P_e^F) = 0,  
\label{eeq}
\end{eqnarray}
which means the balance of electron's kinetic and potential energies in Eq.~(\ref{eeq0}) and degenerate electrons 
occupy energy-levels up to $+m_ec^2$. 
Eqs.~(\ref{en},\ref{eeq0},\ref{eeq}) give the relationships:
\begin{eqnarray}
P_e^F &=& \frac{1}{c}\left[V^2_{\rm coul}(r)+2m_ec^2V_{\rm coul}(r)\right]^{1/2};  
\label{ppeq}\\
n_e(r) &=& \frac {1}{3\pi^2(c\hbar)^3}\left[V^2_{\rm coul}(r)+2m_ec^2V_{\rm coul}(r)\right]^{3/2}.  
\label{en1}
\end{eqnarray}
The Gauss law leads the following Poisson equation and boundary conditions,
\begin{eqnarray}
\Delta V_{\rm coul}(r)= 4\pi \alpha\left[n_p(r)-n_e(r)\right];\quad V_{\rm coul}(\infty)=0,\quad V_{\rm coul}(0)={\rm finite}.
\label{eposs}
\end{eqnarray}
These equations describe a Thomas-Fermi model for neutral nuclear cores, and have numerically solved together with the empirical
formula \cite{ruffinistella80,ruffinistella81} and $\beta$-equilibrium equation \cite{rrx2006}  for the proton number
$N_p$ and mass number $A=N_p+N_n$, where $N_n$ is the neutron number. 

\begin{figure}[th]
\begin{center}
\includegraphics[width=10cm,height=6cm]{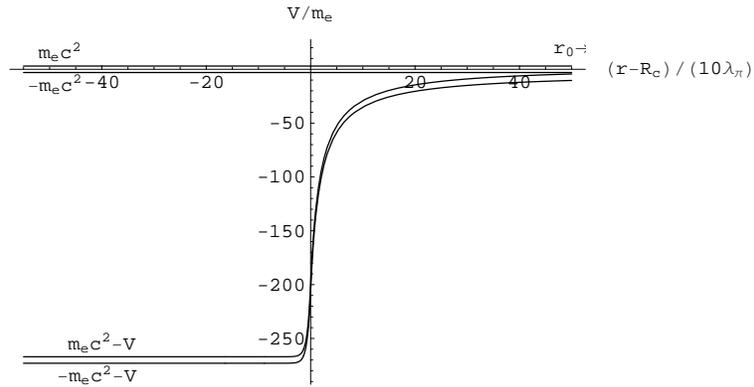}
\end{center}
\caption{Potential energy-gap $\pm m_ec^2-V_{\rm coul}(r)$ and electron mass-gap $\pm m_ec^2$ in the unit of $m_ec^2$ 
are plotted as a function of $(r-R_c)/(10\lambda_\pi)$. The potential depth inside core ($r<R_c$) 
is about pion mass $m_\pi c^2\gg m_ec^2$ and potential energy-gap and electron mass-gap are indicated. The radius $r_0$ where
electron distribution $n_e(r_0)$ vanishes in super charged core case is indicated as $r_0-$, since it is out of plotting range.}%
\label{potential}%
\end{figure}

\vskip0.2cm
\noindent{\it Equilibrium of electron distribution in super charged cores}\hskip0.3cm 
In Ref.~\cite{muller75,migdal76}, assuming that super charged cores of proton density (\ref{pn}) are bare, 
electrons (positrons) produced by vacuum polarization fall (fly) into cores (infinity), one studied the equilibrium of 
electron distribution when vacuum polarization process stop. When the proton density is about nuclear density,
super charged core creates a negative Coulomb potential well $-V_{\rm coul}(r)$, 
whose depth is much more profound than $-m_ec^2$ (see Fig.~[\ref{potential}]), production of electron-positron pairs take places, and 
electrons bound by the core and screen down its charge. Since the phase space of negative energy-levels $\epsilon(p)$
\begin{eqnarray}
\epsilon(p)= [(pc)^2+m_e^2c^4]^{1/2} -V_{\rm coul}(r), 
\label{sefep}
\end{eqnarray}
below $-m_ec^2$ for accommodating electrons is limited, vacuum polarization process completely stops when electrons fully occupy 
all negative energy-levels 
up to $-m_ec^2$, even electric field is still critical. 
Therefore an equilibrium of degenerate electron distribution is expected when the following condition is satisfied, 
\begin{eqnarray}
\epsilon(p)= [(pc)^2+m_e^2c^4]^{1/2} -V_{\rm coul}(r) = -m_ec^2, \quad p=P_e^F,
\label{sefep1}
\end{eqnarray}
and Fermi-energy  
\begin{eqnarray}
{\mathcal E}_e(P_e^F) =\epsilon(P_e^F)- m_ec^2 =-2 m_ec^2 , 
\label{efep}
\end{eqnarray}
which is rather different from Eq.~(\ref{eeq}). 
This equilibrium condition (\ref{efep}) leads to electron's Fermi-momentum and number-density (\ref{en}),
\begin{eqnarray}
P_e^F &=& \frac{1}{c}\left[V^2_{\rm coul}(r)-2m_ec^2V_{\rm coul}(r)\right]^{1/2};  
\label{ppeqp}\\
n_e(r) &=& \frac {1}{3\pi^2(c\hbar)^3}\left[V^2_{\rm coul}(r)-2m_ec^2V_{\rm coul}(r)\right]^{3/2}.  
\label{en1p}
\end{eqnarray}
which have a different sign contracting to Eqs.~(\ref{ppeq},\ref{en1}). Eq.~(\ref{eposs}) remains the same. 
However, contracting to the neutrality 
condition $N_e=N_p$ and $n_e(r)|_{r\rightarrow\infty}\rightarrow 0$ in the case of neutral cores, 
the total number of electrons is given by
\begin{eqnarray}
N^{\rm ion}_e=\int_0^{r_0}  4\pi r^2 drn_e(r) < N_p,
\label{golbalnp1p}
\end{eqnarray}
where $r_0$ is the finite radius at which electron distribution $n_e(r)$ (\ref{en1p}) 
vanishes: $n_e(r_0)=0$ , i.e., $V_{\rm coul}(r_0)=2m_ec^2$, and $n_e(r)\equiv 0$ for the range $r>r_0$. 
$N^{\rm ion}< N_p$ indicates that such configuration is not neutral.
These equations describe a 
Thomas-Fermi model for super charged cores, and have numerically \cite{muller75} and 
analytically \cite{migdal76} solved with assumption $N_p=A/2$.

\vskip0.2cm
\noindent{\it Ultra-relativistic solution}\hskip0.3cm
In analytical approach \cite{migdal76,rrx2007}, the ultra-relativistic approximation is adopted for 
$V_{\rm coul}(r)\gg 2m_ec^2$, the term $2m_ec^2V_{\rm coul}(r)$ in Eqs.~(\ref{ppeq},\ref{en1},\ref{ppeqp},\ref{en1p}) 
is neglected. It turns out that approximated Thomas-Fermi equations are the same for both cases of neutral and charged cores,
and solution $V_{\rm coul}(r)=\hbar c(3\pi^2n_p)^{1/3}\phi(x)$,
\begin{equation}
\phi(x) = \left\{\begin{array}{ll} 1-3\left[1+2^{-1/2}\sinh(3.44-\sqrt{3}x)\right]^{-1}, &  {\rm for}\quad x<0, \\
\frac {\sqrt{2}}{(x+1.89)}, & {\rm for}\quad x>0,
\end{array}\right\},
\label{popovs2}
\end{equation}
where $x=2(\pi/3)^{1/6}\alpha^{1/2} n_p^{1/3} (r-R_c)\sim 0.1  (r-R_c)/\lambda_\pi$ and the pion Compton length
$\lambda_\pi=\hbar/(m_\pi c)$. 
At the core center $r=0(x\rightarrow -\infty)$, $V_{\rm coul}(0)=\hbar c(3\pi^2n_p)^{1/3}\sim m_\pi c^2$. 
On core surface $r=R_c(x=0)$, $V_{\rm coul}(R_c)=3/4V_{\rm coul}(0)\gg m_ec^2$, indicating that the ultra-relativistic 
approximation is applicable for $r\lesssim R_c$. This approximation breaks down at $r\gtrsim r_0 $. Clearly, it is impossible 
to determine the value $r_0$ out of ultra-relativistically approximated equation, 
and full Thomas-Fermi equation (\ref{eposs}) with source terms 
Eq.~(\ref{en1}) for the neutral case, and Eq.~(\ref{en1p}) for the charged case have to be solved.

For $r<r_0$ where $V_{\rm coul}(r)> 2m_ec^2$, we treat the term $2m_ec^2V_{\rm coul}(r)$ in Eqs.~(\ref{en1},\ref{en1p}) as a small 
correction term, and find the following inequality is always true
\begin{equation}
n_e^{\rm neutral}(r)  > n_e^{\rm charged}(r), \quad r < r_0,
\label{ineq}
\end{equation}
where $n_e^{\rm neutral}(r)$ and $n_e^{\rm charged}(r)$ stand for electron densities of 
neutral and super charged cores. 
For the range $r>r_0$, $n_e^{\rm charged}(r)\equiv 0$ in the case of super charged core, 
while $n_e^{\rm neutral}(r)\rightarrow 0$ in the case of neutral core, which should be 
calculated in non-relativistic approximation: the term $V^2_{\rm coul}(r)$ in Eq.~(\ref{en1}) is neglected.
 
In conclusion, the physical scenarios and Thomas-Fermi equations of neutral and super charged cores are slightly different.
When the proton density $n_p$ of cores is about nuclear density, ultra-relativistic approximation applies for
the Coulomb potential energy $V_{\rm coul}(r)\gg m_e c^2$ in $0<r<r_0$ and $r_0 > R_c$, and approximate equations and solutions 
for electron distributions inside and around cores are the same. As relativistic regime $r \sim r_0$ and non-relativistic 
regime $r> r_0$ (only applied to neutral case) are approached, 
solutions in two cases are somewhat different, and need direct integrations.   

\end{document}